\documentclass[aps,pra,twocolumn,superscriptaddress,floatfix,longbibliography]{revtex4-2}

\usepackage[utf8]{inputenc}
\usepackage[english]{babel}
\usepackage{mathtools}
\usepackage{physics}
\usepackage{braket}
\usepackage{bbm}
\usepackage{bm}
\usepackage{amsbsy}
\usepackage{amsthm}
\usepackage{amssymb}
\usepackage{amsfonts}
\usepackage{amsmath}
\usepackage{dsfont} 
\usepackage{graphicx} 
\usepackage{hhline}
\usepackage{epsfig}
\usepackage{epstopdf}
\usepackage{dsfont}
\usepackage{multibib}
\usepackage{xcolor}
\usepackage{color}
\usepackage{verbatim}
\usepackage{nomencl}
\usepackage[colorlinks]{hyperref}
\usepackage{float}
\usepackage[normalem]{ulem}
\usepackage{multirow}
\usepackage{makecell}

\addto\captionsenglish{}

\begin{document}
	
	\title{Genuine \textit{k}-partite correlations and entanglement in the ground state of the Dicke model for interacting qubits}

	\author{Ant\^onio C. Louren\c{c}o}
	\email[]{lourenco.antonio.c@gmail.com}
	
	\affiliation{Departamento de F\'isica, Universidade Federal de Santa Catarina, CEP 88040-900, Florian\'opolis, SC, Brazil}
	
	\author{Denis R. Candido}
	\affiliation{Department of Physics and Astronomy, University of Iowa, Iowa City, Iowa 52242, USA}
	
	\author{Eduardo I. Duzzioni}
	\affiliation{Departamento de F\'isica, Universidade Federal de Santa Catarina, CEP 88040-900, Florian\'opolis, SC, Brazil}
	
	\begin{abstract}

          Here, we calculate and study correlations of the Dicke model in the presence of qubit-qubit interaction. Whereas
          the analysis of correlations among its subsystems is essential for the understanding of corresponding critical phenomena and for performing quantum information tasks, the majority of correlation measures are restricted to bipartitions due to the inherent challenges associated with handling multiple partitions. To circunvent this we employ the calculation of Genuine Multipartite Correlations (GMC) based on the invariance of our model under particle permutation. We then quantify the correlations within each subpart of the system, as well as the percentage contribution of each GMC of order $k$, highlighting the many-body behaviors for different regimes of parameters. Additionally, we show that GMC signal both first- and second-order quantum phase transitions present in the model. Furthermore, as GMC encompasses both classical and quantum correlations, we employ Quantum Fisher Information (QFI) to detect genuine multipartite entanglement. Ultimately, we map the Dicke model with interacting qubits to spin in solids interacting with a quantum field of magnons, thus demonstrating a potential experimental realization of this model.
           
	\end{abstract}
	
	\maketitle

    \section{Introduction}

    Correlations describe properties and characteristics of a composed system that cannot be obtained from its counterparts. Whether classical or quantum, correlations are used and employed extensively in different areas of physics (e.g., in statistical mechanics, quantum mechanics, and condensed matter physics) to understand and characterize the behavior of their systems.   
    Most recently, quantum correlations (e.g., entanglement) have also shown to be important for quantum information science, as they are a key ingredient for several quantum information protocols such as Shor’s factorization algorithm for cryptography breaking \cite{shor1994}, quantum teleportation \cite{horodecki2009}, quantum error correction \cite{Scott2004}, and metrology \cite{toth2012}.
    Moreover, they also serve as resources for quantum batteries \cite{shi2022} and thermal machines capable of reversing heat flow \cite{micadei2019}.

    The study of quantum correlations in many-body systems using entanglement \cite{Osterloh2002,lambert_2004,oliveira_2006b,hannukainen2018}, quantum discord \cite{Sarandy2009,Werlang2010}, and coherence \cite{Li2016,Wang2021b}, has been important for the understanding and characterization of quantum phase transitions and critical phenomena\cite{horodecki2009}. 
     For instance, entanglement measures such as concurrence \cite{Osterloh2002,osborne2002,wang2012}, entanglement of formation \cite{lambert_2004}, entanglement entropy \cite{latorre2005}, global multipartite entanglement \cite{Oliveira_2006a,oliveira_2006b,Oliveira2008}, and entanglement negativity \cite{hannukainen2018}, signal Quantum Phase Transitions (QPTs) through the presence of peaks in the parameter space. Despite being a present general feature, some measures occasionally fail to detect the phase transition~\cite{Liu2010}.
      
    Quantum correlations have been studied in different models, e.g., Ising model \cite{Verstraete2004,Grimmett2008,Giampaolo_2014,Krutitsky2017,schneider2021}, Lipkin-Meshkov-Glick model \cite{latorre2005,Cui2008,Bao2021,lourenco2020}, Dicke superradiance \cite{santos2016,calegari2020,lohof2023}, and the Dicke model \cite{schneider2002,Rossatto2020,soldati_2021,Boneberg2022}, to detect QPT or to describe the types of correlations present in the models. Of current relevance, the Dicke model is important in the fields of quantum optics and condensed matter physics \cite{HEPP_1973,Garraway_2011}. It was created to describe the interaction between light and matter in optical cavities, but it is generally represented as an ensemble of $N$ spins $1/2$ particles (two-level systems) interacting with a single mode of a ``cavity''. In equilibrium, it presents a second-order QPT from a normal to a superradiant phase \cite{HEPP_1973,Garraway_2011}. This model has been widely studied in the literature and has several variations and generalizations that account for different types of qubit-cavity and qubit-qubit interaction \cite{Bowden1978,Zou2014,Lamata2017,robles_ground_2015,Xu2021,Cordero2019,soldati_2021}. In the presence of qubit-qubit interaction, the corresponding generalized Dicke model presents a series of first-order QPT depending on the number of particles and qubit-qubit coupling strength ~\cite{robles_ground_2015}. Some works have investigated correlations such as entanglement and coherence in this model \cite{schneider2002,Rossatto2020,soldati_2021,Boneberg2022}, with multipartite quantum correlations being studied in the two-mode cavity version of the Dicke model \cite{soldati_2021}.   
    
    Calculating or creating operationally feasible measures of correlation beyond two-part systems is a challenging task due to the exponential increase in the calculation time with the number of subsystems. Nevertheless, Girolami et al. \cite{girolami2017} have developed a measure that efficiently calculates genuine $k$-partite correlations in systems invariant under particle permutation. This measure of genuine multipartite correlations (GMC) is known by capturing second-order quantum phase transitions \cite{lourenco2020,calegari2020,lourenco_2022,lourenco_thesis}. However, the study of GMC \cite{girolami2017} and genuine $k$-partite entanglement is still absent for the generalized Dicke Model accounting for qubit-qubit interaction. Accordingly, here we calculate and study the GMC for the generalized Dicke model as a function of the qubit-cavity and qubit-qubit coupling. Most importantly, we obtain the percentage of GMC for each genuine $k$-partite correlation in the model, showing the contribution of the correlation for the many-body behavior.  Also, we show that GMC do signal both the first- and second-order QPTs in the model.   
    As GMC capture all genuine $k$-partite correlations (classical and quantum) in the system, we advance to understand the nature of these correlations by exploring other measures of quantum correlations, such as quantum Fisher information (QFI) \cite{hyllus2012,pezze2009,toth2012} and global entanglement \cite{Oliveira_2006a,oliveira_2006b}. 
    Furthermore, we provide a novel way of implementing the generalized Dicke model in condensed matter systems by interfacing spin centers in solids~(e.g., Nitrogen-Vacancy (NV) centers in Diamond \cite{Awschalom2018,Schirhagl2014,DOHERTY20131,zvi2023engineering} and di-vacancies in SiC~\cite{doi:10.1126/sciadv.abm5912,Awschalom2018,koehl2011room,Seo2016,doi:10.1126/science.aax9406,PRXQuantum.2.040310}) with magnon modes of a magnetic material~\cite{Candido_2021,Fukami2021,fukami2023magnon}. The use of spin centers to realize and study many-body phenomena has already been explored in the context of critical thermalization~\cite{PhysRevLett.121.023601}, Floquet prethermalization in a long-range spin interacting system~\cite{PhysRevLett.131.130401}, and discrete time-crystals~\cite{Choi2017,doi:10.1126/science.abk0603}. Most recently, Ref.~\cite{losey2022solid} proposed the use of spin centers interacting via dipole-dipole interaction for creating quantum simulators of the spin-1/2 XYZ model. 

    This work is presented as follows: in Sec. \ref{sec:dickemodel} we introduce the generalized Dicke model that includes interaction amongst particles. Sec. \ref{sec.gmc} contains the GMC measure. The results are presented in Sec. \ref{sec:results}, which include witnesses and measures of quantum correlations. A discussion on how the generalized Dicke model can be realized in an experimental setting is made in Sec. \ref{sec:nv-centers} and the conclusions and perspectives are in Sec. \ref{sec:conclusion}.

    \section{Generalized Dicke Model}
    \label{sec:dickemodel}
    The Dicke model \cite{dicke54} was first introduced to depict the interaction between atoms and light \cite{HEPP_1973,Garraway_2011}, although it also effectively describes other types of interaction, e.g., electron-phonon~\cite{Kuzmin1989}, qubit-magnon~\cite{xinwei_2018}, and qubit-resonator~\cite{Yu_2022}. Typically, the model assumes $N$ spin-$1/2$ atomic subsystems (two-level systems or simply qubits) coupled to a one-mode cavity, with corresponding Hamiltonian
    \begin{align}
    \label{eq:hamiltonian.dicke}
	    H_1=\omega_c a^{\dagger}a + \omega_0 S_z + \frac{\lambda}{\sqrt{N}}(a^{\dagger}+a)S_x,
    \end{align}
where $\hbar=1$, $\omega_c$ is the frequency of the cavity mode represented by the creation (annihilation) operator $a^{\dagger}$ ($a$), $\omega_0$ the frequency of the qubits, and $\lambda$ the strength of the qubit-cavity coupling. $S_{\alpha}=\sum_{i=1}^{N} \sigma_{\alpha}^{i}$ with $\alpha=x,y,z$ are the collective spin operators with Pauli matrices $\sigma_{\alpha}^{i}$ and total spin $S=N/2$. Here we define the basis that will be used for the diagonalization of the problem. This will be a product state between the Fock states~\cite{Fock1932} of the cavity, i.e., $a^{\dagger}a \ket{n} = n \ket{n}$ with $n\in \mathbb{N}_0$, and the Dicke states~\cite{dicke54} defined by the collective (total) spin projection along the $z$ direction, i.e., $S_z\ket{S,m_s} = m_s \ket{S,m_s}$, with $m_s \in \{-S, -S+1, \cdots ,S-1, S \}$. 

It is well known that as the qubit-cavity interaction strength ($\lambda$) increases, the system described by Eq.~(\ref{eq:hamiltonian.dicke}) undergoes a second-order phase transition from a normal phase to a superradiant one, resulting in the creation of entanglement between cavity mode and qubits \cite{lambert_2004}. In the superradiant phase, the ground state is two-fold degenerate, while in the normal phase, the ground state $\ket{0}\otimes\ket{S,-S} = \ket{0}\otimes\ket{\downarrow, \downarrow, \cdots, \downarrow}$ is unique and separable. In the presence of longitudinal qubit-qubit interaction, the Hamiltonian for the generalized Dicke model reads \cite{robles_ground_2015}
    \begin{align}
        \label{eq:hamiltonian.int}
	    H_2=\omega_c a^{\dagger}a + \omega_0 S_z + \frac{\eta}{N} S_z^2 + \frac{\lambda}{\sqrt{N}}(a^{\dagger}+a)S_x,
    \end{align}
    where $\eta$ is the interqubit coupling strength. In this model, as the qubit-qubit interaction increases, correlations among the qubits intensify in the normal phase \cite{lambert_2004,robles_ground_2015} and the system experiences a series of first-order phase transitions depending on the number of qubits in the ensemble \cite{robles_ground_2015}. In the limit case $\lambda \rightarrow 0$, Hamiltonian Eq.~\eqref{eq:hamiltonian.int} becomes equivalent to the Lipkin-Meshkov-Glick (LMG) model \cite{Lipkin1965}. 
	
        \section{Genuine $k$-partite correlations}
    \label{sec.gmc}
    
    In this section, we introduce a measure of genuine multipartite correlations (GMC) of order $k$, a tool to measure genuine $k$-partite correlations~\cite{girolami2017}. We emphasize that GMC capture both classical and quantum correlations among the subsystems. Therefore, analysis of this measure of correlations alone does not allow us to understand the nature of these correlations. We will address this point in Sec.~\ref{subsec:QFI_QD_GE}, by explicitly calculating the global entanglement among the subsystems, and by witnessing genuine $k$-partite entanglement through the quantum Fisher information. 

    \subsection{Genuine multipartite correlations}
    \label{sec.gmcA}

    The GMC of order $k$ \cite{girolami2017} are quantified using a distance-based measure, which encompasses both classical and quantum correlations. The GMC have been employed in analyzing various many-body systems \cite{calegari2020,lourenco2020,lourenco_2022}, demonstrating its efficacy in capturing phase transitions and elucidating the hierarchical nature of correlations.
    Before showing how to measure genuine $k$-partite correlations, we need to define product states with genuine $k$-partite correlations. The set of states that have up to $k$ subsystems is defined as $P_{k}\coloneqq \left \{\tau_{N}=\bigotimes_{j=1}^{m}\tau_{k_{j}}, \sum^{m}_{j=1}k_{j}=N, k=\max\{k_{j}\}\right \}$, where $\tau_{k_{j}}$ is the density matrix of a subsystem with $k_{j}$ qubits and $\tau_N$ is the product state with genuine $k$-partite correlations. This set contains all the sets $P_{k'}$ with $k'<k$, such that  $P_{1}\subset P_{2}...\subset P_{N-1}\subset P_{N}$.

    The measure of genuine correlations of order higher than $k$ is defined as the smallest distance between the total density matrix of the system $\rho_N$ and $\tau_N$, namely \cite{girolami2017}
    \begin{equation}
        I^{k \to N}(\rho_N) := \underset{\tau_N\in P_k}{\text{min}} S(\rho_N||\tau_N).
        \label{eq:Shigherkgenuine}
    \end{equation}
    in which $S(\rho||\tau) = \tr(\rho \ln \rho - \rho \ln \tau)$ is the quantum relative entropy. For quantum relative entropy the closest state to $\rho_N$ is the product of its marginals $\rho_{k_j}$, such that $\underset{\tau_N\in P_k}{\text{min}} S(\rho_N||\otimes_{j=1}^{m}\tau_{k_j})=S(\rho_N||\otimes_{j=1}^{m}\rho_{k_j})$. The closest product state to $\rho_N$ for systems invariant under particle permutation is 
    \begin{equation}
        \tau_N=\left(\bigotimes_{i=1}^{\lfloor N/k\rfloor}\rho_k\right)\otimes\rho_{N\mod k},
        \label{eq:productstate}
    \end{equation}
    where $\left \lfloor N/k \right \rfloor $ is the floor function and $\rho_{N\mod k}$ the density matrix of the subsystem with $N\mod k$ qubits.
    Therefore, genuine correlations of order higher than $k$ for a system invariant under particle permutations are calculated by \cite{girolami2017} 
	\begin{equation}
	\label{eq:Dis}
	\begin{split}
	I^{k \rightarrow N}(\rho_{N})=& \left \lfloor N/k \right \rfloor S(\rho_{k})+\\
	&(1-\delta_{N\mod k,0})S(\rho_{N\mod k}) - S(\rho_{N}).
	\end{split}
	\end{equation}
    We can understand the GMC of order $k$ as the difference between the genuine correlations of order higher than $k-1$ and $k$, resulting in the following expression     
    \begin{equation}
	\label{eq:Corr}
	I^{k}(\rho_{N})=I^{k-1 \rightarrow N}(\rho_{N})-I^{k \rightarrow N}(\rho_{N}).
	\end{equation}
     This expression captures only genuine correlations among $k$-subsystems. The total correlations, which returns the difference between the informational content of the totally uncorrelated state $\rho_1^{\otimes N}$ and the state of the system $\rho_N$ is obtained from Eq. (\ref{eq:Shigherkgenuine}) for $k=1$ as
    \begin{equation} \label{eq:totalcorr}
	   I^{1}(\rho_{N}) \equiv N S(\rho_{1}) - S(\rho_{N}).
	\end{equation}
     $\rho_1$ is the density matrix for one subsystem only. The total correlations measure, which is based on generalized mutual information \cite{Avis_2008,yang2009}, can be interpreted as the sum of all GMC of order $k=2,...,N$, $I^{1}(\rho_{N})=\sum_{k=2}^N I^k(\rho_N)$, as can be verified from Eqs. (\ref{eq:Corr}) e (\ref{eq:totalcorr}).

    \begin{figure*}
    \centering
    \includegraphics[width=1.0\textwidth]{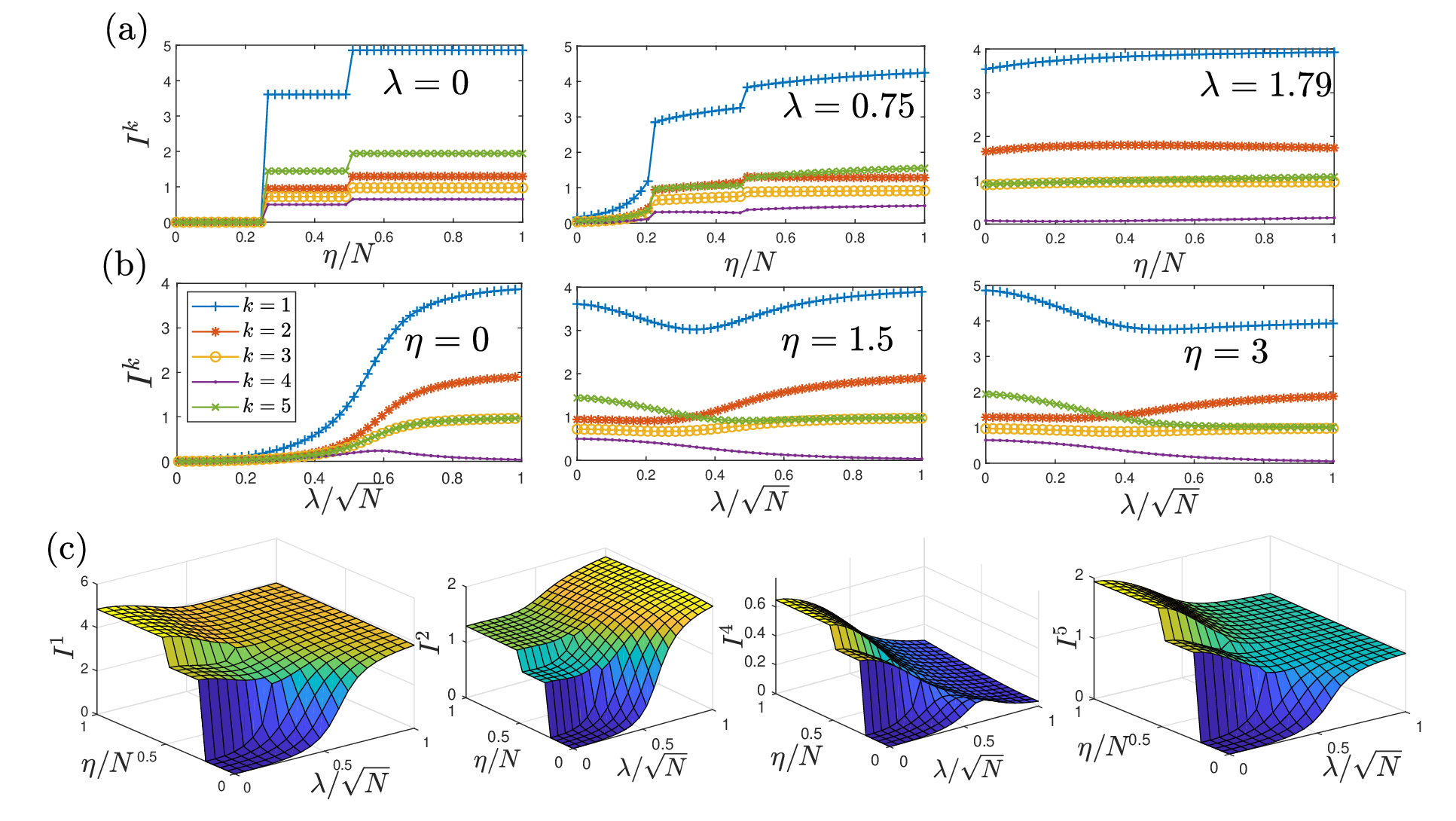}
    \caption{GMC of order $k$, $I^k$, for the ground state of interacting qubits resonant with the cavity mode, $\omega_0=\omega_c=1\xout{.0}$. Considering $N=5$ qubits inside the cavity, in a) we plot $I^k$ as function of interqubit coupling $\eta/N$ for $\lambda = 0$, $0.75$, and $1.79$, while in b) we plot $I^k$ as function of qubit-cavity coupling strength $\lambda/\sqrt{N}$ for $\eta = 0$, $1.5$, and $3$.  The GMC are able of distinguishing both the first-order quantum phase transition (QPT), as pictured in a) for $\lambda=0$, as well as the transition from the normal to the superradiant phases, shown in b) for $\eta=0$. In c), we plot the GMC as function of both $\eta/N$ and $\lambda/\sqrt{N}$ for the ground state of $N=5$ qubits with $\omega_0=\omega_c=1$ and $k=1,2,4,5$. 
    }
    \label{fig:gmc_slices}
    \end{figure*}

    \section{Results} 
    \label{sec:results}
    \subsection{Genuine $k$-partite correlations in the ground state of the generalized Dicke model}   
    
    In order to facilitate the numerical diagonalization of Hamiltonian (\ref{eq:hamiltonian.int}), we performed both a rotation around the $y$-axis, $R_y\left( \pi/2 \right)$, and a displacement of the cavity mode, $D(\beta) = \exp{J_z (\beta a^{\dagger} - \beta^{*} a)}$, with ${\beta=\lambda/\sqrt{N}\omega_c}$ \cite{robles_ground_2015}. After the diagonalization of the rotated Hamiltonian, the corresponding eigenstate is rotated back to its initial basis. In Fig. \ref{fig:gmc_slices}, we present the GMC of order $k$ ($I^k$) [Eq.~(\ref{eq:Corr})] for five qubits under the resonant case $\omega_0=\omega_c$ with different values of qubit-qubit and qubit-cavity couplings. In Fig.~\ref{fig:gmc_slices}(a), we plot $I^k$ as a function of the qubit-qubit coupling $\eta/N$ for three values of qubit-cavity coupling,  $\lambda=0,0.75,1.79$, and for $k=1,2,3,4,5$. Similarly, in Fig.~\ref{fig:gmc_slices}(b) we plot $I^k$ as a function of the qubit-cavity $\lambda/\sqrt{N}$ for $\eta=0,1.5,3$.

    In all panels of Figs.~\ref{fig:gmc_slices}(a) and (b), we observe that GMC of orders $k=2$ and $k=5$ are dominant over correlations of orders $k=3$ and $k=4$. To understand this, we start our discussion by contrasting two limiting cases corresponding to $\lambda=0$ and $\eta=0$. Using $I^1 = I^2 + I^3+I^4+I^5$  presented in  Sec. \ref{sec.gmcA}, we can calculate the percentage contribution of each GMC of order $k$, $100\% \times I_k/I_1$. For $\lambda=0$ and $\eta/N=0.4$ [left panel in Fig.~\ref{fig:gmc_slices}(a)], the GMC of order $k=5$ contribute to $40.0\%$, while $k=2$ with $26.2\%$, $k=3$ with $20.0\%$, and $k=4$ with $13.8\%$. In this regime of parameters, the quantum system exhibits strong correlations and all particles ($k=5$) are significantly more correlated than any smaller subgroup of particles ($k<5$). This is a direct consequence of the qubit-qubit interaction term of our Hamiltonian, $\frac{\eta}{N} S_z^2 = \frac{\eta}{N} \sum_{i,j=1}^N \sigma_z^i \sigma_z^j$, which couples different qubits with same strength, and therefore, do not favor stronger correlation between particular sets of qubits. This interaction Hamiltonian is well known for effectively entangling qubits as a controlled-Z gate~\cite{nielsen00}. Conversely, for $\eta=0$ and $\lambda/\sqrt{N}=0.98$ (left panel in Fig.~\ref{fig:gmc_slices}(b)), the GMC of order $k=2$ contribute to $48.95\%$, while $k=3$ and $k=5$ contribute to $24.99\%$, and $k=4$ to $1.03\%$. In this case, the dominant correlation among the particles is of order $k=2$,  and it is due to the interaction between qubits via $\frac{\lambda}{\sqrt{N}}(a^{\dagger}+a)S_x$. In other words, two particles interact indirectly via the quantum field of the cavity. This type of interaction is widely used for the realization of entanglement between qubits via SWAP or $\sqrt{\textrm{i-SWAP}}$ protocols~\cite{nielsen00,tanamoto2009,rasmussen2020,fukami2023magnon}. Finally, when both types of interactions are present, there is competition between correlations $I^2$ and $I^5$. For instance, for $\eta=1.5$ [middle panel Fig.~\ref{fig:gmc_slices}(b)] and small qubit-cavity coupling $\lambda\lesssim 0.35$, we see a dominance of the correlation among all particles, $I^5$. However, as the qubit-cavity coupling increases, $I^2$ becomes dominant. A similar crossover is also seen in the right panel of Fig.~\ref{fig:gmc_slices}(b) around $\eta/\sqrt{N}\approx 0.4$, and in the middle panel of Fig.~\ref{fig:gmc_slices}(a) around $\eta/N\approx 0.5$. Interestingly, for all the regimes of analyzed parameters, the GMC of order $k=4$ are shown to be lesser than any GMC of order $k\neq4$. Other works have already reported the same finding, attributing this due to finite-size effects \cite{lourenco2020,calegari2020,lourenco_2022,lourenco_thesis}.

     We now move to the discussion of quantum phase transitions.
     In Fig.~\ref{fig:gmc_slices} (a), for $\lambda=0$, the GMC increase discontinuously in steps, which is a characteristic of first-order quantum phase transitions \cite{Sachdev2011}. We observe that there are two first-order phase transitions and the correlation increases with the interaction between qubits. To understand the first-order quantum phase transitions, we analyze the energy spectrum of the corresponding Hamiltonian (\ref{eq:hamiltonian.int}) with $\lambda=0$, i.e.,
    \begin{equation*}
        E(n,m_s) = \omega_c n + \omega_0 m_s + \frac{\eta}{N}m_s^2,
    \end{equation*}
    where $n$ {and $m_s$ are} quantum numbers defined in Sec.~\ref{sec:dickemodel}, representing the number of modes in the cavity and the total spin projection along $z$. Once the cavity energy is positive, its smallest contribution to the energy spectrum is given by $n=0$. Then our analysis is left with the qubits energy term $E_q(m_s) = \omega_0 m_s + \frac{\eta}{N}m_s^2$. For $\eta/N = 0$, the ground state energy is $-\omega_0 N/2$. As the control parameter $\eta$ increases, there will be values of $\eta/N > \omega_0/(N-1) $ for which $E_q(-N/2) > E_q(-N/2+1)$. So, the phase transition occurs at the critical parameter $\eta^c/N= \omega_0 /(N-1)$. This reasoning can extended to other values of $m_s$, provided that the energy spectrum will increase discontinuously for successive values of $m_s$. Then, by imposing the condition $E_q(m_s) > E_q(m_s+1)$, we find that the first-order quantum phase transitions will occur for values of $\eta^c/N = -\omega_0/(2 m_s + 1)$. Also, from $E_q(m_s)$, we notice that positive values of $m_s$ will return only excited energies, which is captured by the above expression, once to obtain positive values of $\eta/N$, $m_s \ge 1/2$. Applying this relation to Fig.~\ref{fig:gmc_slices} (a), in which $\omega_0=1$ and $N=5$, we obtain the first first-order quantum phase transition at $\eta^c/N = 1/4$ and the second one at $\eta^c/N = 1/2$.

     The Hamiltonian eigenstates will determine the behavior of the GMC, and for $\lambda=0$ they read $\{ \ket{n}\otimes \ket{S,m_s} \}$. Therefore, as we increase $\eta$ from zero, the corresponding ground state will change discontinuously between the successive states $\ket{0}\otimes \ket{S,-S}$ and $\ket{0}\otimes \ket{S,-S+1}$ and so on, stopping when $m_s > -1/2$.  The first ground state, $\ket{0}\otimes\ket{S,-S} = \ket{0}\otimes\ket{\downarrow, \downarrow, \cdots, \downarrow}$, is an uncorrelated state, followed by the second one, the correlated $W$-state $\ket{0}\otimes\ket{S,-S+1} =\ket{0}\otimes\frac{1}{\sqrt{N}}(\ket{\uparrow, \downarrow, \downarrow, \cdots, \downarrow} + \ket{\downarrow, \uparrow, \downarrow, \cdots, \downarrow} + \cdots +\ket{\downarrow, \downarrow, \downarrow, \cdots, \uparrow})$ \cite{robles_ground_2015}. Although the concurrence presents a peak during the first transition, as shown in Ref.~\cite{robles_ground_2015}, the values of GMC are maxima only after the second transition. While the concurrence captures only the first first-order QPT, GMC capture all of the first-order QPTs via the divergent behavior of its first-order derivative with respect to $\eta$ at the transition points $\eta^c/N = -\omega_0/(2 m_s + 1)$. Indeed, $\ket{S,m_s}$ are Dicke states, whose GMC have already been discussed in Ref. \cite{calegari2020}. In Fig.~\ref{fig:gmc_slices} (a), the first-order QPT characterized by the discontinuities of the GMC become smoother as we increase the interaction between qubits and cavity $\lambda$,   until it reaches a point where the discontinuities disappear completely [see the plot for $\lambda=1.79$]. This occurs provided that the interacting term between cavity and qubits dominates over the other terms of the energy spectrum. 
    
     In Fig.~\ref{fig:gmc_slices} (b), in particular, for $\eta=0$ the sudden increase of $I^k$ around  
    $\lambda/\sqrt{N}=\sqrt{\omega_c\omega_0}/2=1/2$ ($\omega_c=\omega_0=1$) captures the second-order QPT, which was already studied in Refs. \cite{HEPP_1973,Garraway_2011,robles_ground_2015}. In the normal phase, characterized by $\lambda/\sqrt{N}<1/2$, the GMC decrease and approach zero for $\lambda\rightarrow0$, with corresponding pure and separable state $\ket{0}\otimes\ket{S,-S}$. Similarly to Fig.~\ref{fig:gmc_slices}(a), as we increase the strength of the qubit-qubit interaction $\eta=1.5$ and $3$, the measure of GMC does not detect any phase transition as $\lambda$ changes continuously [See Fig.~\ref{fig:gmc_slices}(b)]. This feature is also present on the entangled web concurrence shown in Ref. \cite{robles_ground_2015}. In Appendix~\ref{appendix}, we discuss QPTs for a higher number of qubits.

    In Fig. \ref{fig:gmc_slices}(c), we plot the genuine $k$-partite correlations as a function of both coupling parameters $\lambda/\sqrt{N}$ and $\eta/N$. The plots not only capture the two QPTs (first and second order) presented in the previous panels but also reveal how correlated the qubits are for a wider range of parameters. For instance, we note that by increasing qubit-cavity coupling $\lambda$ with $\eta/N>1/2$, the qubit state becomes more mixed due to the entanglement between qubits and cavity mode, presenting the $5$-partite correlation weaker and strengthening the $2$-partite correlation. Furthermore, we notice that along the line  $\lambda=0$, the correlation of order $k=5$ is greater than $k=2$. In this situation the qubits state is pure, so increasing the qubit-qubit interaction $\eta$, also increases the GMC, as seen in Fig. \ref{fig:gmc_slices}(a).

    To summarize, the calculation of the GMC shown in Figures \ref{fig:gmc_slices} evidence the presence of both first and second-order QPTs in the Dicke model [Eq.~(\ref{eq:hamiltonian.dicke})], and the evolution of these QPT as we introduce a finite qubit-qubit coupling $\eta$ [Eq.~(\ref{eq:hamiltonian.int})]. However, GMC contains both classical and quantum correlations, making it impossible to draw further conclusions about the character of these correlations.  
   To overcome this, in the following section, we will investigate the existence of quantum correlations among the qubits.

     \subsection{Entanglement in the ground state of the generalized Dicke model}
     \label{subsec:QFI_QD_GE}

      GMC encompass both classical and quantum multipartite correlations among subsystems. Accordingly, in this section we will carry out a careful examination of the presence of entanglement in the interacting Dicke model described by Eq.~(\ref{eq:hamiltonian.int}). Among several quantum correlations measures, entanglement \cite{Braunstein2002,Jozsa2003,shor1994,Scott2004,horodecki2009,toth2012} 
    has shown to be useful as a resource for the realization of many quantum tasks. Hence, we will first apply quantum Fisher information (QFI) \cite{pezze2009,hyllus2012,toth2012} to witness genuine $k$-partite entanglement among the qubits of the generalized Dicke model. Although QFI can witness the presence of genuine $k$-partite entanglement, it does not quantify it. To further complement these results, we apply the measure of global entanglement \cite{Oliveira_2006a,oliveira_2006b}. The calculation of this quantity depends only on the reduced density matrix of the subsystems. 
     \begin{figure}
        \centering
        \includegraphics[width=0.5\textwidth]{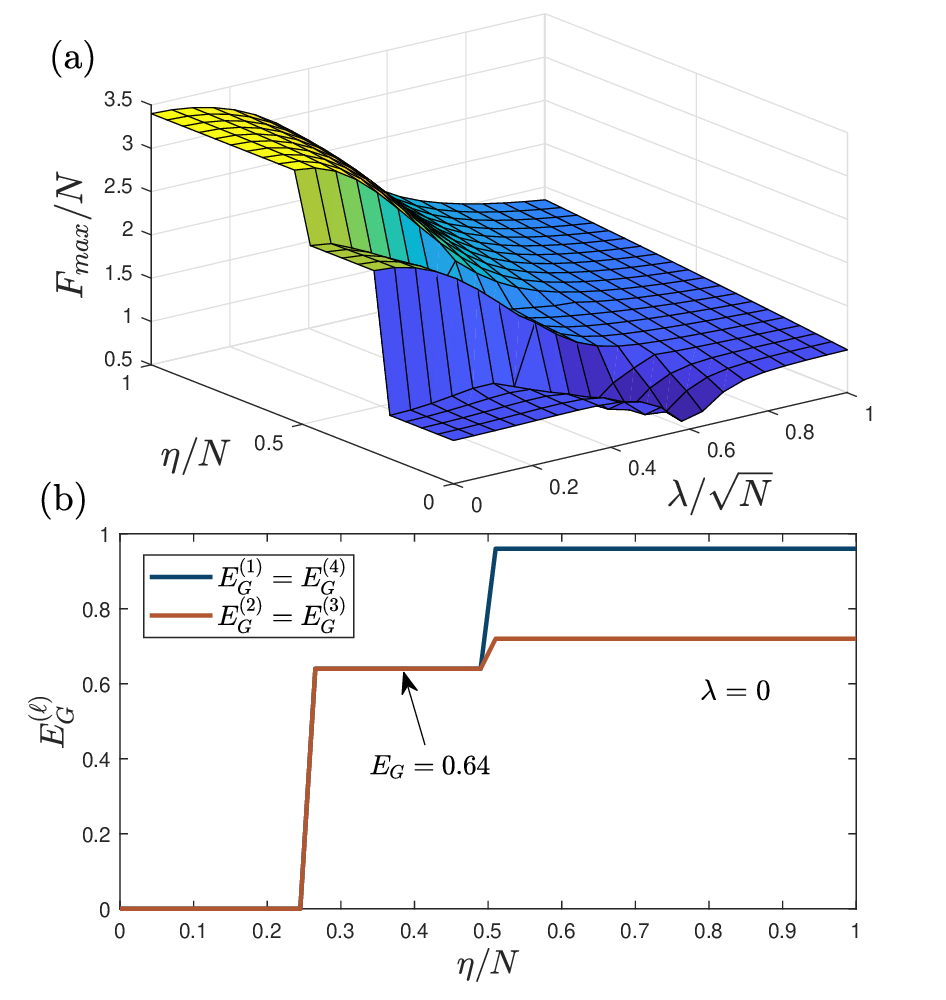}
          \caption{Maximum value of the QFI and generalized global entanglement for the ground state of the generalized Dicke model with $N=5$ qubits interacting resonantly with the cavity, $\omega_0=\omega_c =1$. (a) The $F_{max}/N$ for the whole range of parameters $\eta$ and $\lambda$. The discontinuities along the axis $\eta/N$ show the first-order QPTs, while the deep along the axis $\lambda/\sqrt{N}$ indicates the second-order QPT. (b) We observe the step-like behavior of the generalized global entanglement in the ground state of the system for $\ell=1,2,3,4$ as a function of $\eta/N$ for $\lambda=0$.
    }
        \label{fig:qfi.vs.eta.vs.lambda}
    \end{figure}
     
     \subsubsection{Genuine $k$-partite entanglement}
     In order to witness genuine $k$-partite entanglement among the qubits, we apply the QFI. Specifically, we compute the maximum QFI optimized over the global spin observables $S_\alpha$ ($\alpha=x,y,z$). The maximum QFI for a general state is given by the maximum eigenvalue of the $3\times3$ matrix:
      \begin{equation}
	\label{eq:maxQFI}
	[ \Gamma ]_{\alpha \beta}=2\sum_{i,j}\frac{(p_i-p_j)^2}{p_i+p_j}\bra{j}S_{\alpha}/2\ket{i}\bra{i}S_{\beta}/2\ket{j},   
    \end{equation}
    where $\alpha,\beta=x, y, z$, $p_i+p_j>0$, and $ \rho=\sum_i p_i\ket{i}\bra{i}$ is the spectral decomposition of the state of the quantum system. We denote $F_{\max}$ as the maximum of the QFI. If $F_{\max}(\rho)/N>(k-1)$, the state $\rho$ has genuine $k$-partite entanglement. However, if $F_{\max}(\rho)<N$ nothing can be said, and it might be the case in which the witness $F_{\max}$ just failed to detect the genuine $k$-partite entanglement.

    In the upper panel of Fig. \ref{fig:qfi.vs.eta.vs.lambda}, we plot the maximum of the QFI as a function of $\lambda/\sqrt{N}$ and $\eta/N$ for five qubits. We note that for values where the qubit-cavity coupling is zero ($\lambda=0$), the first-order QPTs are evidenced via the discontinuity of $F_{max}$. In this case, for each excitation in the Dicke state that describes the ground state of the qubits (as we increase $\eta$) more particles get entangled, increasing the degree of genuine $k$-partite entanglement. Moreover, for $\lambda=0$, the $W$-state possesses genuine $3$-partite entanglement, so that the maximum entanglement detected is genuine $4$-partite for the second first-order QPT, represented by the second step in the graph. The GMC show us that for $5$ qubits there are also genuine $5$-partite correlations and all smaller orders of correlations are present (see Fig. \ref{fig:gmc_slices}). For fixed values of $\eta > 0$ and increasing $\lambda$, we observe a strong reduction of the QFI. The genuine $k$-partite entanglement for $\lambda/\sqrt{5}\gtrsim 1/2$ (QPT point) is witnessed only for genuine $2$-partite entanglement.  After crossing the second-order phase transition ($\lambda/\sqrt{5}\gtrsim 1/2$), the reduced state of the qubits is highly mixed because of the entanglement between the qubits and the cavity. Therefore, we understand the suppressing of the higher-partite entanglement as a consequence of QFI being a bad witness for mixed states~\cite{Fiderer2019}. These results highlight and build upon previous findings, demonstrating that, beyond the presence of genuine $k$-partite correlations, genuine $k$-partite entanglement is also present and thus serves as a valuable resource for various quantum tasks.

    \subsubsection{Generalized global entanglement}
     Generalized global entanglement is a measure applicable only for pure states. In summary, it measures through the linear entropy \cite{Oliveira_2006a,oliveira_2006b,Cui2008} the average entanglement between any subsystem of size $\ell$ and the rest of the system \cite{Oliveira_2006a,oliveira_2006b,Cui2008}.   As our Hamiltonian Eq.~(\ref{eq:hamiltonian.int}) only supports pure ground states for $\lambda=0$, we can only use this measure to quantify the entanglement as a function of the qubit-qubit interaction strength. According to the Refs \cite{Oliveira_2006a,oliveira_2006b,Cui2008} for a system invariant under particles permutation the generalized global entanglement is calculated as 
    \begin{align}
        E^{(\ell)}_{G}(\rho)=\frac{d}{d-1}\left[1-\Tr(\rho_\ell^2)\right],
    \end{align}
    where $\ell$ is the number of particles in the reduced density matrix $\rho_{\ell}$ and $d$ is the dimension of the smallest subsystem between the subsystems with $\ell$ particles and its complement with $N-\ell$ particles.  
    
    In the lower panel of the Fig. \ref{fig:qfi.vs.eta.vs.lambda}, we calculate the generalized global entanglement $E^{(\ell)}_{G}$ for $\lambda=0$, $\ell=1,2,3,4$, and $5$ qubits. The average entanglement between qubits increases in discrete steps as we increase the qubit-qubit interaction strength. The step transitions happens at each first-order QPT. In the region with corresponding ground state W-state, the generalized global entanglement is equal to $E^{(1)}_G=4(N-1)/N^2=0.64$ and $E^{(2)}_G=16(N-2)/(3N^2)=0.64$, as shown by Oliveira \textit{et al.} \cite{Oliveira_2006a}. These results are crucial for understanding the extent of generalized global entanglement in the Dicke model, as the Quantum Fisher Information (QFI) only serves as a witness to genuine $k$-partite entanglement without quantifying its magnitude. Additionally, as shown in Ref. \cite{robles_ground_2015}, the system also exhibits pairwise qubit entanglement, evidenced by a nonzero concurrence.

     \subsection{Effective interacting Dicke Hamiltonian with spin centers in solids}
    \label{sec:nv-centers}
    In this section, we show that the Hamiltonian of an ensemble of spin centers (e.g., NV-center) interacting with magnon modes of a magnetic material, can be mapped into the generalized Dicke Hamiltonian with qubit-qubit coupling [Eq.~(\ref{eq:hamiltonian.int})]. Spin centers are defects with spin in solids, and have been shown to be long-lived qubits and very sensitive quantum sensors~\cite{doi:10.1126/sciadv.abm5912,Awschalom2018,Schirhagl2014,DOHERTY20131,koehl2011room,Seo2016,doi:10.1126/science.aax9406,PRXQuantum.2.040310,candido-spin1surface,candido-spin1surface,zvi2023engineering}. NV-centers are amongst the most promising spin centers. Their interaction with magnon modes has been extensively investigated by theoretical and experimental works~\cite{Candido_2021,Fukami2021,fukami2023magnon,main_PhysRevX.3.041023,main_PhysRevLett.121.187204,main_candido2020predicted,main_gonzalez2022towards,main_PhysRevB.105.245310,main_hetenyi2022long,main_PhysRevLett.125.247702}. The corresponding Hamiltonian for magnons interacting with NV-centers was rigorously derived in Refs.~\cite{Candido_2021,Fukami2021,fukami2023magnon}. In the presence of an external magnetic field $B$ along the NV main symmetry axis it reads \cite{Candido_2021,Fukami2021,fukami2023magnon}
    \begin{align}
H_{3} & =\sum_{\mu}\omega_{\mu}b_{\mu}^{\dagger}b_{\mu}+\sum_{i}\frac{\omega_{\textrm{NV}}}{2}\sigma_{\textrm{NV}_{i}}^{z} \nonumber \\
 & +\sum_{i,\mu}g_{\mu}\left(\mathbf{r}_{\textrm{NV}_{i}}\right)\left(b_{\mu}^{\dagger}+b_{\mu}\right)\left({\textbf{\ensuremath{\sigma}}}_{\textrm{NV}_{i}}^{+}+{\textbf{\ensuremath{\sigma}}}_{\textrm{NV}_{i}}^{-}\right). \label{nv-magnon}
\end{align} 
The first term is the magnon Hamiltonian with $\omega_{\mu}$ being the frequency and $b_{\mu}$ $(b_{\mu}^{\dagger})$ the annihilation (creation) operator of the magnons modes $\mu$. The second term represents the NV-center Hamiltonian, where $\omega_{\textrm{NV}}=D_{\textrm{NV}}-\gamma B$ is the frequency splitting between the lowest NV-levels $\left|e\right\rangle$ and $\left|g\right\rangle$, $D_{\textrm{NV}}$ the NV-center zero-field splitting, and $\sigma_{\textrm{NV}}^z=\left|e\right\rangle \left\langle e\right|-\left|g\right\rangle \left\langle g\right|$. The last term describes the interaction between magnon modes $\mu$ and NV-center $i$ (located at $\textbf{r}_{\textrm{NV}_i}$) with corresponding coupling strength $g_\mu \left(\mathbf{r}_{\textrm{NV}_{i}}\right)$, and $\sigma_{\textrm{NV}}^+=\left|e\right\rangle \left\langle g\right|$.

    In order to map the Hamiltonian Eq.~(\ref{nv-magnon}) to the generalized Dicke model Eq.~(\ref{eq:hamiltonian.int}), we need to make a few considerations. First, we use magnon-NV coupling as independent of the position, i.e., $g_\mu \left(\mathbf{r}_{\textrm{NV}_{i}}\right)=g_\mu $. This can be done by implanting the NV-centers in regions where the amplitude of the magnon fringe field is uniform ~\cite{Candido_2021,Fukami2021}. Secondly, we separate the magnons into two different sets: one set containing one single magnon mode $\nu$ with a frequency close to $\omega_{\textrm{NV}}$, i.e., $\omega_\nu\approx\omega_{\textrm{NV}}$ (on-resonant magnon), and the other set with off-resonant magnons.  It is well known that off-resonance quasi-particles mediate the coupling between different spins and qubits~\cite{Candido_2021,Fukami2021,fukami2023magnon,PhysRevB.106.L180406,li2023solid}, e.g., Ruderman–Kittel–Kasuya–Yosida (RKKY) interaction \cite{PhysRev.96.99,10.1143/PTP.16.45,PhysRev.106.893}. Using second order perturbatin theory, the Hamiltonian for qubits coupled via off-resonant magnons reads $\sum_{i,j} g_{\textrm{eff}} \sigma_{\textrm{NV}_{i}}^z\sigma_{\textrm{NV}_{j}}^z/N$~\cite{PhysRevB.106.L180406,li2023solid} with $g_{\textrm{eff}}=|g_{\mu}|^2/\omega_{\mu}$. Furthermore, considering low temperatures such that the excitation of the set of off-resonant magnons modes is small, i.e., $\left\langle b_{\mu}^\dagger b_{\mu}\right\rangle \ll 1 $, Hamiltonian Eq.~(\ref{nv-magnon}) becomes 
    \begin{align}
          H &= \omega_{\nu}b_{\nu}^\dagger b_{\nu} + \sum_i\frac{\omega_{\textrm{NV}}}{2}\sigma_{\textrm{NV}_i}^z \\
          & + \frac{g_{\textrm{eff}}}{N}\sum_{i,j} \sigma_{\textrm{NV}_i}^z\sigma_{\textrm{NV}_j}^z + g_\nu (b_{\nu}^\dagger + b_{\nu})\sum_i\left(\sigma_{\textrm{NV}_{i}}^{+}+\sigma_{\textrm{NV}_{i}}^{-}\right).\nonumber
    \end{align}
    Defining the collective spin operators $S^z=\sum_{i=1}^{N} \sigma_{\textrm{NV}_i}^z$ and $S^x=\sum_{i=1}^{N} \sigma_{\textrm{NV}_i}^x$, we obtain
     \begin{align}
          H = \omega_{\nu}b_{\nu}^\dagger b_{\nu} + \frac{\omega_{\textrm{NV}}}{2} S^z + \frac{g_{\textrm{eff}}}{N}\left(S^z\right)^2+ g_\nu (b_{\nu}^\dagger + b_{\nu})S^x.\nonumber
    \end{align} 
    
    By comparing the equation above with Eq.~(\ref{eq:hamiltonian.int}), we identify $\lambda/\sqrt{N}=g_\nu$, $\eta=g_{\textrm{eff}}$ and  $\omega_0=\omega_\textrm{NV}/2$.
    
   \section{Conclusions} \label{sec:conclusion}
  
 We analyzed the Genuine Multipartite Correlations (GMC) among qubits in the ground state of the generalized Dicke model. Our study identified the percentage contribution of each GMC of order $k$ to the total correlations and established a connection between the Hamiltonian interaction term and the order of the correlation. The findings reveal that all orders of GMC, as well as genuine $k$-partite entanglement (as evidenced by the quantum Fisher information), effectively detect both first- and second-order quantum phase transitions in the model. Additionally, we propose a novel system for implementing the generalized Dicke model using NV centers interacting with magnons in low-loss magnetic materials.

	\begin{acknowledgments}
		The authors acknowledge financial support from
		the Brazilian funding agencies Coordenação
		de Aperfeiçoamento de Pessoal de Nível Superior
		(CAPES), Conselho Nacional de Desenvolvimento
		Científico e Tecnológico (CNPq), Fundação de Amparo à Pesquisa e       Inovação do Estado de Santa Catarina (FAPESC), and Instituto
		Nacional de Ciência e Tecnologia de Informação
		Quântica (CNPq INCT-IQ (465469/2014-0)). EID thanks to CNPq grant number $409673/2022-6$.  D.R.C. acknowledges support from the U. S. Department of Energy, Office of Science BES Award No. {DE-SC002339}.
		
	\end{acknowledgments}

	\begin{appendix}
        
	\appendix
 
	\begin{widetext}
    \section{Quantum Phase Transitions}
    \label{appendix}
     To further study the first order QPTs in the ground state of the system, in Fig. \ref{fig:gmc_QPT}(a) we plot GMC of orders $k=1,...,10$ as a function of the interqubit coupling $\eta/N$ for $\lambda=0$ and $N=20$ qubits. First, we notice that the GMC capture all the first-order QPTs. Secondly, the range of $\eta$ in which the $W$ state is the ground state (i.e., between first and second first-order QPTs) is smaller compared to the one for $N=5$, shown in Fig.~\ref{fig:gmc_slices}(a), which also corroborates the results for the concurrence studied in Ref. \cite{robles_ground_2015}.
      \begin{figure}[H]
    \centering
    \includegraphics[width=0.96\textwidth]{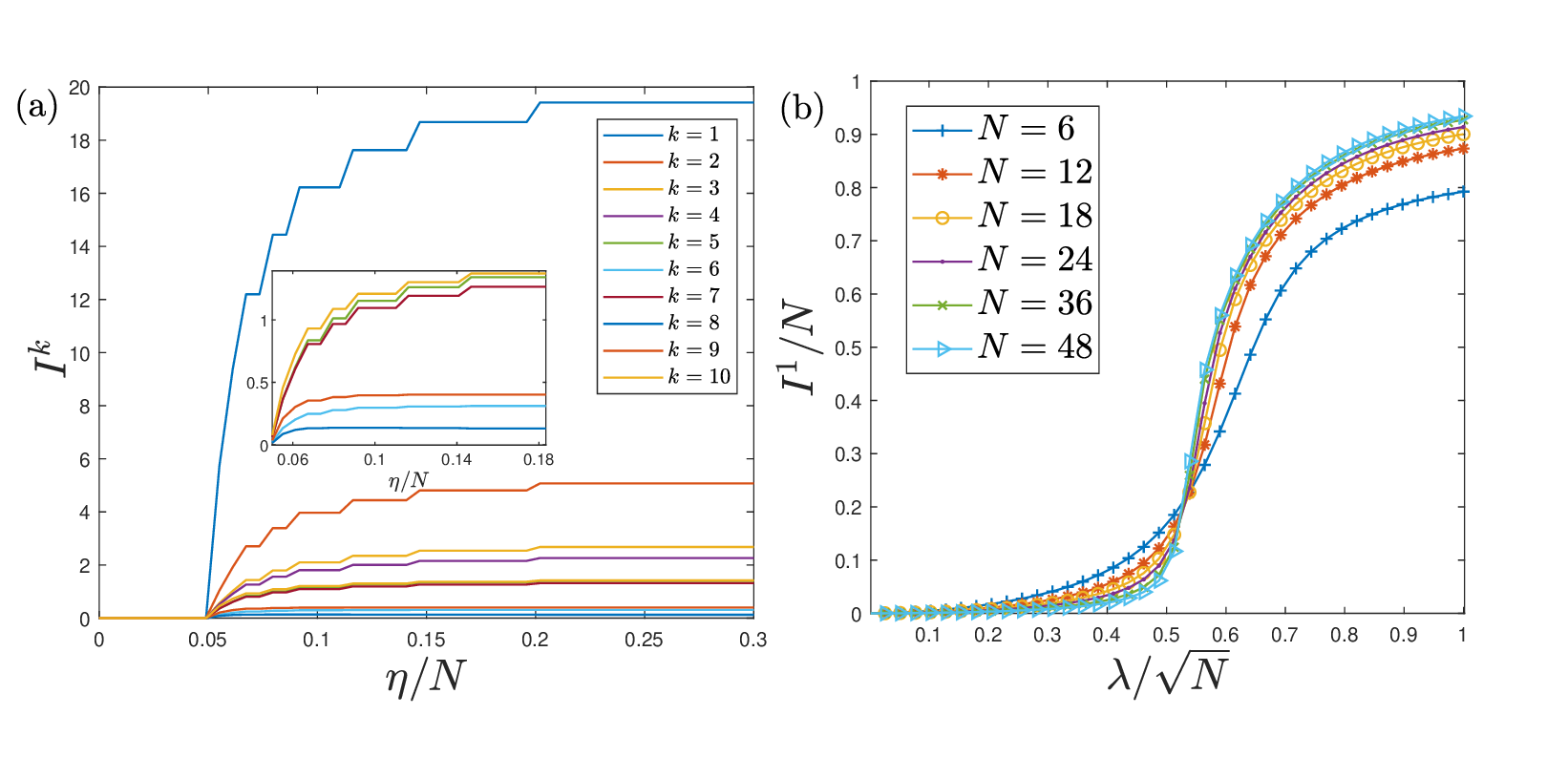}
    \caption{(a) Plot of the GMC $I^k$ as a function of interqubit coupling $\eta$ for the ground state of $N=20$ qubits resonant with the cavity mode  $(\omega_0=\omega_c=1.0)$ with $\lambda=0$. The first-order QPTs are all captured by the GMC, even though for small values of genuine $k$-partite correlations the steps become less evident, as noticed in the inset. (b) The plot of the total correlations $I^1$ as a function of the qubit-cavity coupling $\lambda$ for $\eta=0$ and increasing values of the number of qubits, $N$. The second-order QPT, which happens in the thermodynamic limit $N\rightarrow\infty$, is captured by the abrupt increase of total correlations between the normal and superradiant phase at $\lambda_c/\sqrt{N}=1/2$.}
    \label{fig:gmc_QPT}
    \end{figure}
    In a second-order QPT in the boundary time crystal \cite{Iemini2018}, it was shown that GMC are extensive with the number of particles in the paramagnetic phase and subextensive in the ferromagnetic one \cite{lourenco_2022}. The result for the Lipkin-Meshkov-Glick Hamiltonian \cite{lourenco2020} pointed to this type of behavior as well. To confirm the extensivity and subextensivity of the phases of the non-interacting Dicke model [Eq.~(\ref{eq:hamiltonian.dicke})], in Fig.~\ref{fig:gmc_QPT}{(b)} we plot the total correlations $I^1$ for an increasing number of qubits $N$. The plots verify that the total correlations are subextensive in the normal phase and extensive in the superradiant phase of the Dicke model, respectively. Although we can only draw this conclusion in the thermodynamic limit ($N\rightarrow\infty$), the maximum number of particles that our computation can reach is $N=48$. Furthermore, augmenting the number of qubits, the inflection point moves towards $\lambda/\sqrt{N}=1/2$, which is known as the critical point of the second-order QPT of the Dicke model. Furthermore, Fig. \ref{fig:gmc_slices}(b) indicates the disappearance of this second-order QPT for $\eta >0$. For this reason, we do not analyze the extensiveness property of the generalized Dicke model. As a final remark, we emphasize that genuine $k$-partite correlation serves as a reliable indicator for both types of quantum phase transitions in the Dicke model, unlike other correlation measures, such as concurrence  \cite{robles_ground_2015}.
    \end{widetext}
    \end{appendix}

	\bibliography{references}

\end{document}